\documentclass[10pt]{JHEP3}

\usepackage{amsmath,amssymb,graphicx,caption}

\newcommand\fverb{\setbox\pippobox=\hbox\bgroup\verb}
\newcommand\fverbdo{\egroup\medskip\noindent%
                        \fbox{\unhbox\pippobox}\ }
\newcommand\fverbit{\egroup\item[\fbox{\unhbox\pippobox}]}
\newbox\pippobox

\title{Hypercharged Conformally Sequestered Gauge Mediation}

\author{
Hae Young Cho  \\
  FPRD and Department of
Physics and Astronomy, Seoul National University, Seoul 151-747,
Korea \\
  E-mail: \email{hycho@phya.snu.ac.kr} }

\maketitle

\abstract{The $B\mu/\mu$ solution in GMSB via the hidden sector
dynamics is simple and natural. However, it has some obstacles to be
physical. To circumvent this situation, we introduce the visible and
the hidden branes, each of which has its own $U(1)$ symmetry, in a
five dimensional setup. In the bulk we allow Chern-Simons coupling
between the visible and the hidden $U(1)$s which gives an
enhancement of the mass of bino. If this gives a considerable
contribution to the mass of bino, we can get a proper radiative
electro weak symmetry breaking with the boundary conditions, in
which $B\mu$ and squared scalar masses are suppressed at the scale,
where the hidden sector is integrated out.}


\begin{document}

\def\lsl{ l \hspace{-0.45 em}/}
\def\ksl{ k \hspace{-0.45 em}/}
\def\qsl{ q \hspace{-0.45 em}/}
\def\psl{ p \hspace{-0.45 em}/}
\def\ppsl{ p' \hspace{-0.70 em}/}
\def\dsl{ \partial \hspace{-0.45 em}/}
\def\Dsl{ D \hspace{-0.55 em}/}
\def\matrix{ \left(\begin{array} \end{array} \right) }
\def\frsqotw{\frac{1}{\sqrt2}}
\def\frsqoth{\frac{1}{\sqrt3}}
\def\frsqtth{\sqrt{\frac{2}{3}}}
\def\frsqo{\frac{\omega}{\sqrt{3}}}
\def\frsqoo{\frac{\omega^2}{\sqrt{3}}}

\def\hf{\textstyle{\frac12~}}
\def\hff{\textstyle{\frac13~}}
\def\hfg{\textstyle{\frac23~}}
\def\DQ{$\Delta Q$}

\def\Qem{Q_{\rm em}}

\thispagestyle{empty}

\baselineskip 0.6cm


\section{Introduction}
Sequestering is a mechanism that can suppress the amplitudes of the
unwanted operators. This is usually used when we want to suppress
the tree level flavor changing neutral current (FCNC) in the
mediation mechanism, where the gravity may have a significant
contribution. It can be understood in the geometrical sense via the
string theory \cite{sequ}. Sequestering gives a number of
interesting phenomenological features so it is worth studying.

There are a variety of mediation mechanisms in MSSM. Among them the
most famous ones, which are free from FCNC problem are gauge
mediated supersymmetry breaking (GMSB) and anomaly mediated
supersymmetry breaking (AMSB). In GMSB, generating an electro weak
scale $\mu$ is not serious by itself however, the requirement for
the low energy electro-weak symmetry breaking (EWSB) makes a very
unnatural situation \cite{Dvali:1996cu}. Among a number of possible
explanations to ameliorate this
\cite{muGMSB,Roy:2007nz,Murayama:2007ge}, the sequestering idea
suggests a very simple solution \cite{Roy:2007nz,Murayama:2007ge}.
By the way, once the information of the hidden sector
renormalization effect is imposed, we have an additional effect: the
gravitino has an enhanced mass. This aspect also makes the
sequestering idea interesting in GMSB. The idea to solve $B\mu/\mu$
problem in GMSB via the conformal sequestering is clear and simple.
However, the simplest form appears not to have physical case, i.e.
it does not seem to provide a physical solution to a proper EWSB
\cite{Cho:2008fr,Asano:2008qc}. It is because the boundary
conditions given at an intermediate scale have relatively small
scalar squared masses including higgs. When we follow along the MSSM
RG equation, it is hard to satisfy the EWSB conditions. In other
words, the parameter space where we have a proper EWSB is not
compatible with the boundary conditions at the intermediate scale.
Therefore some modification to the simplest case is necessary.
\begin{figure}[t]
\begin{center}
\includegraphics[width=8cm]{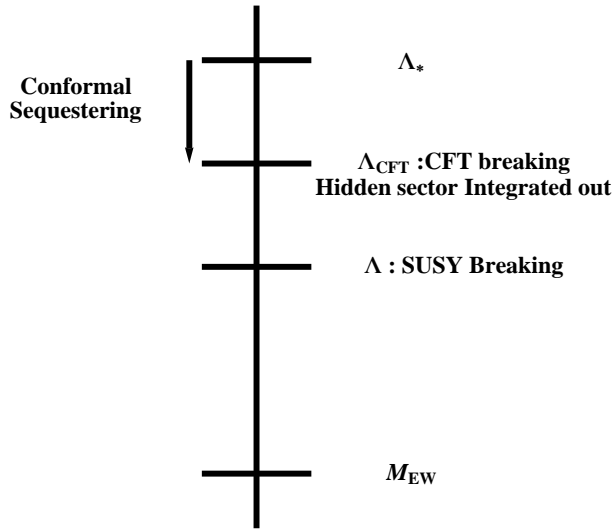}
\end{center}
\caption{Scale and the dynamics of
Ref.\cite{Roy:2007nz,Murayama:2007ge}}\label{CzA}
\end{figure}
Here we consider the 5 dimensional setup. We introduce two 3 branes:
one is where the visible sector resides and the other is for hidden
sector. We introduce $U(1)_h$ at the hidden brane and a five
dimensional Chern-Simons coupling in the bulk. As an effect of five
dimensional Chern-Simons term, the mass of bino on the visible brane
is enhanced. This is the idea in \cite{Dermisek:2007qi} to solve the
tachyonic slepton problem in a pure AMSB setup. Just like
\cite{Dermisek:2007qi}, in this setup, all the scalars charged under
$U(1)_Y$ get the radiative correction by the mass of bino. With a
numerical study we find a physical solution, i.e. get a proper EWSB.

This paper is organized as follows. In section 2, we discuss on the
sequestering in GMSB. In section 3, we take the viewpoint of the
five dimensional setup and introduce a Chern-Simons term in the bulk
to find a physical solution. In section 4, we discuss the
phenomenological implication. Finally we make a conclusion.

\section{Sequestering in Gauge Mediation}
Strictly speaking, sequestering in GMSB is not a necessary condition
to circumvent FCNC problem because supersymmetry breaking occurs at
a rather low energy scale so that undesirable gravity contribution
is negligible. The role of sequestering in GMSB, however, appears to
be interesting and attractive because of its unique
feature\cite{bmura}. In GMSB there is a problem known as $B\mu/\mu$
problem, and it is recently suggested that if we consider the
sequestering effect in GMSB, then we can solve $B\mu/\mu$ problem
\cite{Roy:2007nz,Murayama:2007ge}. Here we will briefly review the
idea of sequestering as a solution to $B\mu/\mu$ problem. In the
supersymmetry conserving part, $\mu$ is the unique dimensionful
parameter so that it is necessary to link it with supersymmetry
breaking to ensure the low energy supersymmetry. In GMSB, it is
possible to introduce the superpotential as
\begin{equation}\label{hsp}
W=\lambda X\textbf{10}\cdot
\bar{\textbf{10}}+\xi_dH_d\textbf{10}\cdot\textbf{10}+\xi_uH_u\bar{\textbf{10}}\cdot\bar{\textbf{10}},
\end{equation}
where $\textbf{10},\bar{\textbf{10}}$ are messenger fields, $X$
denotes the field which breaks supersymmetry and, $\lambda$ and
$\xi_{u,d}$ are $\mathcal{O}(1)$ appropriate dimensionless
couplings. After integrating out the massive messenger fields, we
get
\begin{equation}\label{ob}
\begin{split}
\mu\sim&\frac{\xi_{u}\xi_{u}}{16\pi^2}
H_1H_2\left(\frac{M^{\dagger}}{M}\frac{F^{\dagger}}{M^{\dagger}}\right)
=\frac{\xi_{u}\xi_{d}}{16\pi^2}\Lambda
\left[1+\mathcal{O}(\frac{F^2}{M_{mess}^4})\right],\\
B_{\mu}\sim&\frac{\xi_{u}\xi_{d}}{16\pi^2}
H_1H_2\left(\frac{M^{\dagger}}{M}(\frac{FF^{\dagger}}{MM^{\dagger}})\right)
=\frac{\xi_{u}\xi_{d}}{16\pi^2}\Lambda^2
\left[1+\mathcal{O}(\frac{F^2}{M_{mess}^4})\right]=\Lambda\mu,\\
A_{H_{1,2}}\sim&\frac{\partial \log{Z_{H_{u,d}}}}{\partial
\log{X}}=\frac{-\frac{|\xi_{u,d}|^2}{16\pi^2}}{1-\frac{|\xi_{u,d}|^2}{16\pi^2}\log{XX^{\dagger}}}\frac{F}{M}\sim
-\frac{|\xi_{u,d}|^2}{16\pi^2}\frac{F}{M},\\
m_{H_{u,d}}^2\sim&\frac{\partial^2 \log{Z_{H_{u,d}}}}{\partial
\log{X}
\partial \log{X^{\dagger}}}=-\left(
\frac{|\xi_{u,d}|^2}{16\pi^2}\right)^2\frac{1}{(1-\frac{|\xi_{u,d}|^2}{16\pi^2}\log{XX^{\dagger}})^2}\frac{FF^{\dagger}}{MM^{\dagger}}\sim
-\left(
\frac{|\xi_{u,d}|^2}{16\pi^2}\right)^2\frac{FF^{\dagger}}{MM^{\dagger}},
\end{split}
\end{equation}
Here we see $B\mu=\Lambda\mu$, which is undesirable for the
phenomenological requirement. This is the $B\mu/\mu$ problem in
GMSB. Here we see that the relation between $A$ and $\mu$, i.e.
$A_{u}A_{d}=|\mu|^2$. Not to conflict to the perturbation, the
coupling $\xi_{u,d}$ should not be large. For a convenience, we take
$\xi_u=\xi_d$, but later we will consider a general case in range,
where the perturbation of the messenger coupling is guaranteed.

The basic idea to solve this problem, which we concern, is using the
1PI effect on the propagator of $X$ in the strongly interacting
hidden sector. As a result, the operators which are proportional to
$XX^{\dagger}$ are suppressed relative to those which are
proportional to either $X$ or $X^{\dagger}$. The operators for
supersymmetry breaking masses are generated at the original
messenger scale just in the case of usual GMSB setup. As we go down
to low energy, the theory goes through a conformal window. At an
intermediate scale $\Lambda_{CFT}$, the conformal symmetry is broken
and the supersymmetry breaking operators get their values, which are
affected by the hidden sector RG effect described above. This
imposes that $B\mu$ can be made of $\mathcal{O}(\mu^2)$ or smaller,
therefore $B\mu/\mu$ problem in GMSB can be solved with a simple
assumption. In addition to this, there can be another effect, coming
from the anomalous dimension of $X$,\footnote{This anomalous
dimension should be considered independently i.e. the effect which
suppresses the operators containing $XX^{\dagger}$ is another. To
make this difference clear, see \cite{Murayama:2007ge,Asano:2008qc}}
which makes the masses of ordinary superpartners suppressed relative
to the mass of the gravitino. In other words, the amount that the
gravitino feels by supersymmetry breaking is not the same as the
others.

Here we want to make it clear whether this mechanism spoils the nice
feature of GMSB in solving FCNC problem or not. The gravitational
contribution to soft breaking parameters appears to be proportional
to the mass of gravitino, which can be a source of the FCNC problem.
As denoted above, the gravitino mass is enhanced by the hidden
sector RG effect and the gauge contribution suppressed by a factor
of anomalous dimensions. Here we can see that a tension between FCNC
and $B\mu/\mu$ problem. As denoted in \cite{Murayama:2007ge} if the
gravity contribution is enhanced, then one of the virtues of GMSB is
lost. Since the problematic contribution to FCNC appears in soft
scalar mass squares, we should be careful about this inequality,
\begin{equation}
m_{gravity}^2\sim\frac{F^{\dagger}F}{M_{P}^2}=m^{2}_{3/2}<m_{gauge}^2.
\end{equation}
By the language appearing in \cite{Murayama:2007ge}, the constraint
on the mass parameters is given by
\begin{equation}\label{grvi}
\frac{(16\pi^2)^2}{\lambda^2}\frac{M_{mess}^2}{M_p^2}\left(\frac{\Lambda_{\ast}}{\Lambda_{CFT}}\right)^{2\gamma_{X}}<\mathcal{O}(1),
\end{equation}
where $\lambda$ is the coupling given in (\ref{hsp}),
$\Lambda_{\ast}$ is a scale where the hidden sector gets conformal
and $\gamma_{X}$ is an anomalous dimension of $X$. This can be
easily satisfied if
$\left(\frac{\Lambda_{\ast}}{\Lambda_{CFT}}\right)^{2\gamma_{X}}$ is
not large.\\

Now we investigate the low energy physics with a numerical tool.
\begin{equation}\label{bc}
\begin{split}
&m_i\sim\frac{\alpha_i}{4\pi}\Lambda \text{   (i=1,2,3)}, \qquad
A_{H_{u,d}}\sim-\mu, \\
&m^{2}_\phi\sim0, \qquad m^{2}_{H_{u,d}}+\mu^2\sim 0
\end{split}
\end{equation}\footnote{There is a confusion on this boundary
conditions \cite{Craig:2008vs,Perez:2008ng}. Since $\mu$ and $A$ are
generated via supersymmetry breaking $F$ terms, a survey on the
effective Kahler potential shows that higgs mass is given as
$m_h^2+|\mu|^2$. By the hidden sector RG effects higgs masses is
vanishes like the other scalar.} With these boundary conditions, we
use \texttt{softsusy} to investigate the low energy spectrum
\cite{Allanach:2001kg}. The idea is very simple and natural however,
it appears to have an unnatural situation in RG improved studies
\cite{Cho:2008fr,Asano:2008qc}. In the MSSM RG equations, we see the
scalar masses are determined by the gaugino contribution, the
trilinear terms and their masses. In this analysis, we just use the
value of $\mu$, which is given by the low energy requirement, that
is we can not handle $\mu$ and $A_{u,d}$ at $\Lambda_{CFT}$. This
makes the analytic approach to this problem difficult. Since
$A_{u,d}$ grow via mainly the gluino contribution and give
considerable radiative correction to higgs mass because of large
yukawa couplings especially top yukawa at small $\tan{\beta}$
region. From the RG equation of MSSM, we see that gauginos give
positive contribution and, $A_{u,d}$ and scalar masses give negative
contribution to higgs masses as we go down to the low energy scale.
Since $A_{u,d}$ get larger than wino and bino, the higgs masses can
get negative. On the other hand, because of $A_l$, the lightest stau
gets negative squared mass, which is not favored. If we restrict
$B\mu$ to a positive definite quantity at the low energy scale in
order not to have a tadpole problem, $B\mu/\mu$, which is under the
control of only gauginos and $A_{u,d,l}$ can grow large enough to
threaten the stability of the higgs potential in some parameter
space. Unfortunately $B\mu$ is also a given quantity in this
analysis, we check whether it is from the boundary condition, which
we assume at $\Lambda_{CFT}$. And the result was the boundary
condition is not compatible with the low energy physics
\cite{Cho:2008fr}. It also appears that it is hard to make things
better if $\Lambda$ is larger than about $200$TeV. Though it is not
easily seen, if we have a large $\Lambda$, then all the soft terms
get the radiative corrections from rather massive gauginos, of which
masses are proportional to $\Lambda$ at $\Lambda_{CFT}$. As denoted
above, the gluino gives large correction to $A_{u,d}$ so that $B\mu$
gets negative at electro weak scale. In \cite{Asano:2008qc}, the
authors try to evade this problem by introducing additional
messenger masses by an adjoint chiral multiplet, which is supposed
to break the grand unified theory. As a result, they are free from
that unnatural situation however, the scalar masses appear small so
that the experimental bound might be dangerous. In the next section,
we suggest another way out.

\section{Introducing a Chern-Simons Term in the Bulk}
Recently, it is found that a hidden $U(1)_h$ has interesting
phenomena in the low energy physics. Among a number of solution to
solve the tachyonic slepton problem, there is a study where hidden
$U(1)_h$ takes an important role
\cite{Dermisek:2007qi,Verlinde:2007qk}. There they consider two
three branes, which have their own $U(1)$ gauge theory: $U(1)_v$ for
the visible brane and $U(1)_h$ for the hidden brane. In the five
dimensional bulk we introduce a Chern Simons coupling $\int
C_{p-1}\wedge tr F$, where $C_{p-1}$ is Ramond-Ramond $p-1$ form in
the bulk. As a result, two $U(1)$ gauge fields get mixed, so that
there can be two linearly independent combinations. One of them
remains massless, which will be $U(1)_Y$, and the other gets
massive. By this mechanism, bino can get an additional contribution,
and the scalar partners of fermions as well as the higgs get
radiative correction
\begin{equation}
\delta
m_{i}^2=-\frac{3}{10\pi^2}g_{1}^2Y_{i}^2M_{1}^2\log{\frac{\mu}{M}}.
\end{equation}
Therefore, the boundary condition at UV scale can be changed. Here
we want to do the same job in the conformally sequestered GMSB
setup.
\begin{equation}\label{bcmodi}
\begin{split}
&m_i\sim\frac{\alpha_i}{4\pi}\Lambda, \text{   (i=2,3)} \qquad
A_{H_{u,d}}\sim-\mu, \\
&m^{2}_\phi\sim0, \qquad m^{2}_{H_{u,d}}+\mu^2\sim0, \\
&m_1\sim\frac{\alpha_1}{4\pi}\Lambda+\text{(enhancement by CS
interaction: $\tilde{M}$)}
\end{split}
\end{equation}
Then we use a package \texttt{softsusy} again to check whether our
modification works. For simplicity, we will consider $\Lambda_{CFT}$
to be close to the original messenger scale because we do not want
to consider the visible sector RG effect. This also help us to
consider the more suppressed case than a naive expectation of
$16\pi^2$, because it can minimize the possible visible sector
contribution. The boundary condition (\ref{bcmodi}) is used at the
effective messenger scale, and below that scale, it is a good
approximation to use MSSM RG. We set the effective messenger scale
as $10^{12}$GeV, varying $\tan{\beta}$ and $\Lambda$. The important
ingredient is the mass of bino correction from CS interaction. This
depends on our choice, and we assume that it is order of
$\frac{\mu}{\tilde{M}}\sim\mathcal{O}(1)$\footnote{Unlike the
original idea, the gravitino mass is not an order parameter in this
case. So we make a use of $\mu$, which is made for a proper EWSB.}.
One question may arise on the additional CP phase. Though we assume
that supersymmetry breaking is up to a single field $X$, there can
be a misalignment between other gauginos and bino so that there can
exist the additional CP phase. For simplicity, however, we assume
there is no additional CP phase. With these boundary conditions, we
run \texttt{softsusy}.
\begin{figure}[h]
\begin{minipage}{68mm}
\includegraphics[width=0.9\textwidth]{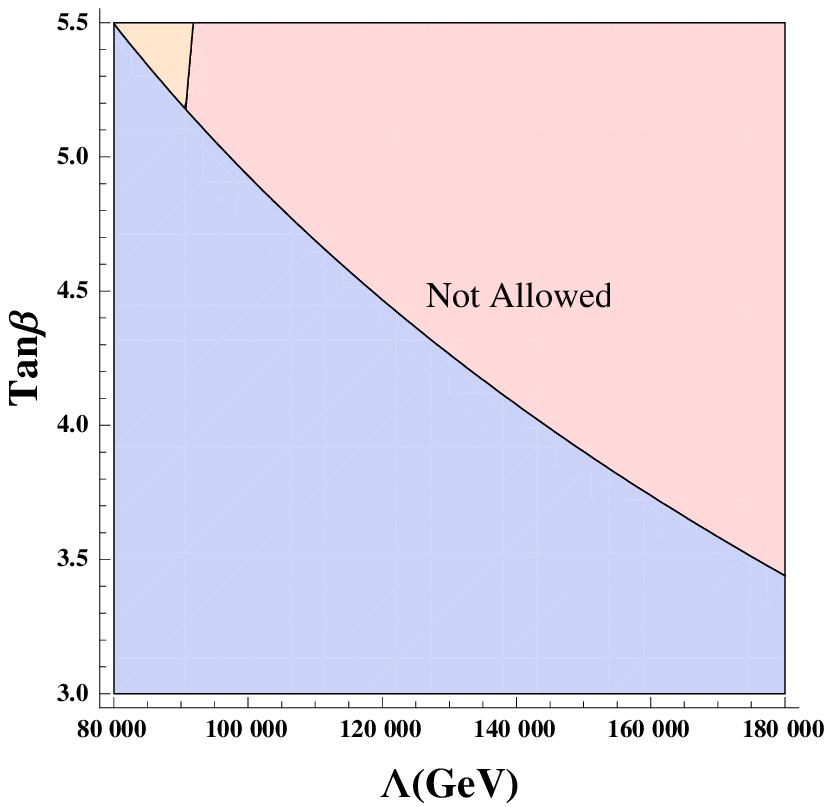}
\end{minipage}
\hfil
\begin{minipage}{68mm}
\includegraphics[width=0.9\textwidth]{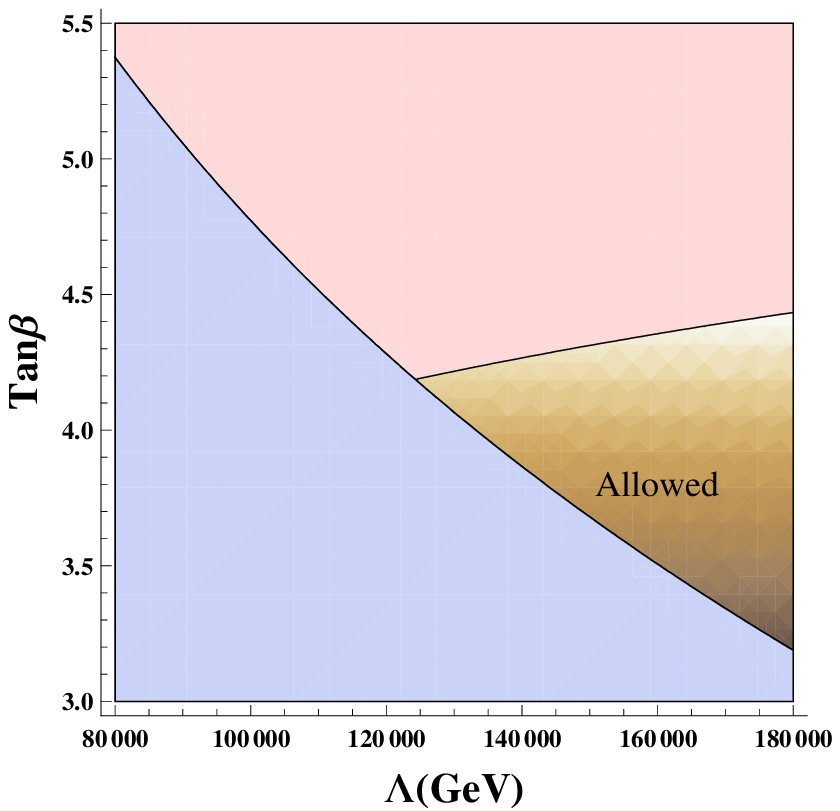}
\end{minipage}
\caption{We choose $\Lambda_{CFT}=10^{12}$GeV and set sign of $\mu$
to be positive. The blue region is excluded by the mass bound for
lightest higgs, the red by the inconsistency and the orange by the
stau mass bound. The left panel is the result of $\tilde{M}=0$ and
the right one is the result of $\tilde{M}=5\mu$. In the right, the
colored with gradient is allowed region, and the brighter is the
better.}\label{1}
\end{figure}\footnote{There can be an error of $\pm3$GeV theoretically in the spectrum calculating
packages, so we allow $3$GeV
 difference in the higgs mass
\cite{Allanach:2004rh}.}

 In the right panel of Fig.~(\ref{1}) we see that there exists
parameter space where the low energy EWSB requirements and the
consistency of mechanism are satisfied. The region can be changed,
when we allow correction to the mass of bino, i.e. this can be tuned
by appropriate $\tilde{M}$ and suppression factor, which is given as
boundary conditions at $\Lambda_{CFT}$.
\begin{figure}[b]
  \begin{center}
  \includegraphics[width=7cm]{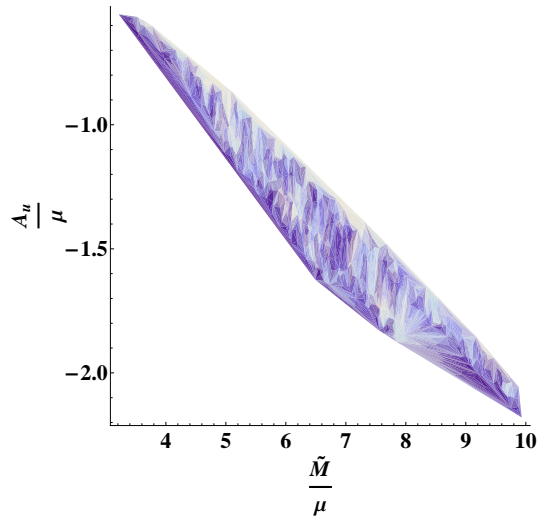}\\
  \caption{The region which satisfies the boundary conditions at the electro weak scale and $\Lambda_{CFT}$. It is projected to the $\frac{\tilde{M}}{\mu}$ and $\frac{A_{u}}{\mu}$ plane.}\label{2}
    \end{center}
\end{figure}
To see the effect of $\tilde{M}$, here we do a simple numerical
analysis. In addition to that, we consider a general case for
$\xi_{u,d}$. We do a numerical analysis varying
$\frac{\tilde{M}}{\mu}$ and $\frac{A_{u}}{\mu}$, then search for the
valid region, which satisfies the boundary conditions. In
Fig.~(\ref{2}) we collected the valid points which satisfy at least
the low energy requirements and
$\left|\frac{B\mu}{\mu^2}\right|<0.01$. The bright part is the
allowed region in the previous analysis. Here we see that the
variation on $\tilde{M}$ and $\xi_u$ is restricted by the boundary
conditions. If $\Lambda_{CFT}$ is as low as $10^8$GeV, the result is
slightly deformed. But the conclusion that we can find the parameter
region, where the low energy requirements are satisfied with
appropriate $\frac{\tilde{M}}{\mu}$ and $\frac{A_{u}}{\mu}$, is not
changed.

\section{Phenomenology}
In the parameter space which passes the low energy requirement, we
pick a typical point, and analyze it. As we discussed in the
previous section, here we have three dimensionful parameters:
$\Lambda$, $\Lambda_{CFT}$ and $\tilde{M}$. In addition to them, we
have 2 dimensionless parameter: $\tan\beta$ and $\frac{A_{u}}{\mu}$.
These are under the control of the low energy constraints. Here we
investigate the case
\begin{equation}\label{bcpoint}
\begin{split}
&\Lambda = 1.52089\times 10^5 GeV, \qquad \Lambda_{CFT}=10^{12}GeV,
\qquad \tilde{M}=5 \mu,\qquad \tan \beta=4.0, \qquad A_u=-\mu.
\end{split}
\end{equation}

\begin{tabular}[c]{|c|c|c|c|c|c|c|c|c|c|c|}
  \hline
   $\chi^0_1$&$\chi_1^{+}$&$\tilde{t}_1$&$\tilde{b}_1$&$\tilde{u}_L$&$\tilde{c}_L$&$\tilde{d}_L$&$\tilde{s}_L$&$\tilde{g}$&$h_0$&$H$ \\
  \hline
   378.3 & 393.3 & 785.7& 802.0 & 9897.3 & 987.3 & 997.7 & 997.7&  1119.3 & 111.98 & 1094.54\\
  \hline
   $\tilde{d}_R$&$\tilde{s}_R$&$\tilde{b}_2$&$\tilde{\nu}_{\tau}$&$\tilde{\nu}_e$&$\tilde{\nu}_{\mu}$&$\tilde{\tau}_1$&$\tilde{\mu}_L$&$\tilde{e}_L$&$\chi^0_2$&$A$\\
  \hline
   1127.3 & 1127.3 & 1131.9 & 1085.7 & 1086.6 & 1086.6 & 1093.7 & 1093.4 & 1094.8 & 1216.7& 1093.83\\
    \hline
   $\chi^{+}_2$&$\chi^{0}_3$&$\tilde{t}_2$&$\tilde{u}_{R}$&$\tilde{c}_R$&$\tilde{\tau}_{2}$&$\tilde{e}_R$&$\tilde{\mu}_R$&$\chi^0_4$&$H^{\pm}$&$\mu$\\
  \hline
   1221.5 & 1124.2 & 1454.6 & 1644.8 & 1644.8 & 2127.0 & 2128.0 & 2128.0& 4052.3 & 1097.12& 1224.4\\
  \hline
\end{tabular}
\\
\\
Note that the lightest superpartner except gravitino is wino, and
the lightest scalar partner is the lightest stop. This can be
understood easily because we give a correction to the mass of bino.
Adding this correction makes $\mu$ as large as few TeV therefore the
absolute value of trilinear coupling grows as $\mu$ at
$\Lambda_{CFT}$. Here we assume that the trilinear coupling
generated by the messenger higgs coupling, so that the values of the
trilinear couplings $A_{u,d}$ are obtained with an ambiguity
$\frac{A_{u}}{A_{d}}$. If we look the RG flow of the trilinear
coupling, $A_{u}$ and $A_{d}$ grow monotonically as we go down to
the low energy scale. Moreover small $\tan \beta$ means that top
yukawa coupling is large relatively to others therefore, large off
diagonal term in the stop mass matrix makes stop lighter than any
other scalar partners. Now we consider some physical constraints,
especially the decay rate of the rare process $B\rightarrow
X_s\gamma$ and the anomalous magnetic moment of muon. Here is our
result at the point given in (\ref{bcpoint}), using
\texttt{microOmegas}\cite{momega}.
\begin{equation}\label{cons}
\begin{split}
&(g-2)_{\mu}=5.60 \times 10^{-11},\\
&Br(B\rightarrow X_s\gamma)=3.69\times 10^{-4}.\\
\end{split}
\end{equation}
The anomalous magnetic moment of muon is reported to be $\triangle
a_{\mu}=(30.2\pm8.7)\times 10^{-10}$ in \cite{g-2}, and the rare
process $B\rightarrow X_s\gamma$ is reported as $Br(B\rightarrow
X_{s}\gamma)=(355\pm24^{+9}_{-10}\pm3)\times 10^{-6}$ in \cite{bsg}.
The LSP is definitely gravitino, though the gravitino receive the
correction form the anomalous dimension of $X$. It is because the
gravity contribution is not negligible if very large correction is
applied to the gravitino mass as shown in (\ref{grvi}). As we accept
large $\Lambda$, the masses are increased. If we take the anomalous
dimension effect on gravitino into consideration, the gravitino
remains to be the LSP.

\section{Conclusion}
The idea using the hidden sector dynamics to solve $B\mu/\mu$
problem in GMSB is quite simple and natural. However, it needs some
modification to get a physical solution. There might be a number of
ways to circumvent this situation. We propose a simple method via
five dimensional setup, which includes a bulk CS interaction between
two $U(1)$ gauge fields. Within a reasonable parameter region, we
get a physical solution. The typical region appears to have small
$\tan\beta$, which is near 4.

\begin{acknowledgments}
We thank to H.D. Kim and D.Y. Kim for useful discussion. This work
was supported by the Korea Research Foundation grants
(KRF-2008-313-C00162).
\end{acknowledgments}



\begin{thebibliography}{99}


\def\apj#1#2#3{Astrophys.\ J.\ {\bf #1} #2 (#3) }
\def\ijmp#1#2#3{Int.\ J.\ Mod.\ Phys.\ {\bf #1} #2 (#3) }
\def\mpl#1#2#3{Mod.\ Phys.\ Lett.\ {\bf A#1} #2 (#3)}
\def\nat#1#2#3{Nature\ {\bf #1} #2 (#3) }
\def\npb#1#2#3{Nucl.\ Phys.\ {\bf B#1} #2 (#3) }
\def\plb#1#2#3{Phys.\ Lett.\ {\bf B#1} #2 (#3) }
\def\prd#1#2#3{Phys.\ Rev.\ {\bf D#1} #2 (#3)}
\def\pr#1#2#3{Phys.\ Rev.\ {\bf #1} #2 (#3) }
\def\prl#1#2#3{Phys.\ Rev.\ Lett.\ {\bf #1} #2 (#3)}
\def\prp#1#2#3{Phys.\ Rep.\ {\bf #1} #2 (#3)}
\def\sjnp#1#2#3{Sov.\ J.\ Nucl.\ Phys.\ {\bf #1} #2 (#3)}
\def\zp#1#2#3{Z.\ Phys.\ {\bf #1} #2 (#3)}
\def\jhep#1#2#3{JHEP\ {\bf #1} #2 (#3) }
\def\epjc#1#2#3{Euro. Phys. J.\ {\bf C#1} #2 (#3)}
\def\rmp#1#2#3{Rev. Mod. Phys.\ {\bf #1} #2 (#3) }
\def\prgth#1#2#3{Prog. Theor. Phys.\ {\bf #1} #2 (#3)}

\bibitem{sequ}
  S.~Kachru, L.~McAllister and R.~Sundrum,
  JHEP {\bf 0710}, 013 (2007)
  [arXiv:hep-th/0703105].

  M.~Schmaltz and R.~Sundrum,
  JHEP {\bf 0611}, 011 (2006)
  [arXiv:hep-th/0608051].
  M.~Ibe, K.~I.~Izawa, Y.~Nakayama, Y.~Shinbara and T.~Yanagida,
  Phys.\ Rev.\  D {\bf 73}, 015004 (2006)
  [arXiv:hep-ph/0506023].

  A.~G.~Cohen, T.~S.~Roy and M.~Schmaltz,
  JHEP {\bf 0702}, 027 (2007)
  [arXiv:hep-ph/0612100].

\bibitem{Dvali:1996cu}
  G.~R.~Dvali, G.~F.~Giudice and A.~Pomarol,
  Nucl.\ Phys.\  B {\bf 478}, 31 (1996)
  [arXiv:hep-ph/9603238].

\bibitem{muGMSB}
  K.~Choi and H.~D.~Kim,
  \prd {61}{015010}{2000}
  [arXiv:hep-ph/9906363]
  ; A.~Delgado, G.~F.~Giudice and P.~Slavich,   
  \plb {653}{424}{2007}
  [arXiv:0706.3873 [hep-ph]]
  ; L.~J.~Hall, Y.~Nomura and A.~Pierce,
  \plb{538}{359}{2002}
  [arXiv:hep-ph/0204062]
  ; K.~S.~Babu and Y.~Mimura,
  [arXiv:hep-ph/0101046]
  ; T.~Yanagida,
  \plb{400}{109}{1997}
  [arXiv:hep-ph/9701394].
  ; G.~F.~Giudice, H.~D.~Kim and
  R.~Rattazzi,   
  [arXiv:0711.4448 [hep-ph]]
  ; M.~Dine, Y.~Nir and Y.~Shirman,
  \prd{55}{1501}{1997} [arXiv:hep-ph/9607397].
  C.~Csaki, A.~Falkowski, Y.~Nomura and T.~Volansky,
  arXiv:0809.4492 [hep-ph].



\bibitem{Roy:2007nz}
  T.~S.~Roy and M.~Schmaltz,
  Phys.\ Rev.\  D {\bf 77}, 095008 (2008)
  [arXiv:0708.3593 [hep-ph]].

\bibitem{Murayama:2007ge}
  H.~Murayama, Y.~Nomura and D.~Poland,
  Phys.\ Rev.\  D {\bf 77}, 015005 (2008)
  [arXiv:0709.0775 [hep-ph]].


\bibitem{Cho:2008fr}
  H.~Y.~Cho,
  JHEP {\bf 0807}, 069 (2008)
  [arXiv:0802.1145 [hep-ph]].


\bibitem{Craig:2008vs}
  N.~J.~Craig and D.~R.~Green,
  arXiv:0808.1097 [hep-ph].

\bibitem{Perez:2008ng}
  G.~Perez, T.~S.~Roy and M.~Schmaltz,
  arXiv:0811.3206 [hep-ph].



\bibitem{Asano:2008qc}
  M.~Asano, J.~Hisano, T.~Okada and S.~Sugiyama,
  arXiv:0810.4606 [hep-ph].

\bibitem{bmura}
  H.~S.~Goh, S.~P.~Ng and N.~Okada,
  JHEP {\bf 0601}, 147 (2006)
  [arXiv:hep-ph/0511301].

  M.~Ibe, K.~I.~Izawa, Y.~Nakayama, Y.~Shinbara and T.~Yanagida,
  Phys.\ Rev.\  D {\bf 73}, 015004 (2006)
  [arXiv:hep-ph/0506023].


  M.~Ibe, Y.~Nakayama and T.~T.~Yanagida,
  Phys.\ Lett.\  B {\bf 649}, 292 (2007)
  [arXiv:hep-ph/0703110].
  S.~Shirai, F.~Takahashi, T.~T.~Yanagida and K.~Yonekura,
  Phys.\ Rev.\  D {\bf 78}, 075003 (2008)
  [arXiv:0808.0848 [hep-ph]].



\bibitem{Dermisek:2007qi}
  R.~Dermisek, H.~Verlinde and L.~T.~Wang,
  Phys.\ Rev.\ Lett.\  {\bf 100}, 131804 (2008)
  [arXiv:0711.3211 [hep-ph]].
\bibitem{Verlinde:2007qk}
  H.~Verlinde, L.~T.~Wang, M.~Wijnholt and I.~Yavin,
  JHEP {\bf 0802}, 082 (2008)
  [arXiv:0711.3214 [hep-th]];




\bibitem{Allanach:2001kg}
  B.~C.~Allanach,
  Comput.\ Phys.\ Commun.\  {\bf 143}, 305 (2002)
  [arXiv:hep-ph/0104145].
\bibitem{Allanach:2004rh}
  B.~C.~Allanach, A.~Djouadi, J.~L.~Kneur, W.~Porod and P.~Slavich,
  JHEP {\bf 0409}, 044 (2004)
  [arXiv:hep-ph/0406166].

\bibitem{momega}
  G.~Belanger, F.~Boudjema, A.~Pukhov and A.~Semenov,
  Comput.\ Phys.\ Commun.\  {\bf 149}, 103 (2002)
  [arXiv:hep-ph/0112278].
  G.~Belanger, F.~Boudjema, A.~Pukhov and A.~Semenov,
  Comput.\ Phys.\ Commun.\  {\bf 174}, 577 (2006)
  [arXiv:hep-ph/0405253].
  G.~Belanger, F.~Boudjema, A.~Pukhov and A.~Semenov,
  Comput.\ Phys.\ Commun.\  {\bf 176}, 367 (2007)
  [arXiv:hep-ph/0607059].


\bibitem{g-2}
  G.~W.~Bennett {\it et al.}  [Muon G-2 Collaboration],
  Phys.\ Rev.\  D {\bf 73}, 072003 (2006)
  [arXiv:hep-ex/0602035].
  K.~Hagiwara, A.~D.~Martin, D.~Nomura and T.~Teubner,
  Phys.\ Lett.\  B {\bf 649}, 173 (2007)
  [arXiv:hep-ph/0611102].
  M.~Davier,
  Nucl.\ Phys.\ Proc.\ Suppl.\  {\bf 169}, 288 (2007)
  [arXiv:hep-ph/0701163].

\bibitem{bsg}
  C.~Amsler {\it et al.}  [Particle Data Group],
  Phys.\ Lett.\  B {\bf 667}, 1 (2008).
  E.~Barberio {\it et al.}  [Heavy Flavor Averaging Group (HFAG)
                  Collaboration],
  arXiv:0704.3575 [hep-ex].



\end{thebibliography}
\end{document}